\documentclass[aip,rsi,floatfix,superscriptaddress,showpacs,amssymb,preprint]{revtex4-1}
\usepackage{amsmath}
\usepackage{mathrsfs}
\usepackage{graphicx}% Include figure files
\usepackage{dcolumn}% Align table columns on decimal point
\usepackage{bm}% bold math
\usepackage{euscript}
\usepackage[dvips]{color}
\usepackage{array}
\def\bra#1{{\langle#1|}}
\def\cg(#1,#2)(#3,#4)(#5,#6){\bra{#1,#2,#3,#4}#5,#6\rangle}
\def\threej(#1,#2)(#3,#4)(#5,#6){\begin{pmatrix}#1&#3&#5\\#2&#4&#6\end{pmatrix}}
\def\sixj(#1,#2,#3)(#4,#5,#6){\begin{Bmatrix}#1&#2&#3\\#4&#5&#6\end{Bmatrix}}
\def\ninej(#1,#2,#3)(#4,#5,#6)(#7,#8,#9){\begin{Bmatrix}#1&#2&#3\\#4&#5&#6\\#7&#8&#9\end{Bmatrix}}
\begin{document}
\setcounter{section}{0}
\pagenumbering{arabic}
\setcounter{footnote}{0}

% ********************
\title{Small-sized Dichroic Atomic Vapor Laser Lock (DAVLL)}
\author{Changmin Lee\footnote{These authors have contributed equally to this work.}\footnote{cmpius@berkeley.edu}}
\affiliation{Department of Physics, University of California, Berkeley, California 94720-7300}
\author{G. Z. Iwata\footnotemark[1]\footnote{g.iwata@berkeley.edu}}
\affiliation{Department of Physics, University of California, Berkeley, California 94720-7300}
\author{E. Corsini}
\affiliation{Department of Physics, University of California, Berkeley, California 94720-7300}
\author{J. M. Higbie}
\affiliation{Department of Physics and Astronomy, Bucknell University, Lewisburg, Pennsylvania, 17837}
\author{S. Knappe}
\affiliation{National Institute of Standards and Technology, Boulder, Colorado 80305-3322}
\author{M.~P.~Ledbetter}
\affiliation{Department of Physics, University of California, Berkeley, California 94720-7300}
\author{D. Budker\footnote{budker@berkeley.edu}}
\affiliation{Department of Physics, University of California, Berkeley, California 94720-7300}
\affiliation{Nuclear Science Division, Lawrence Berkeley National Laboratory, Berkeley, California 94720}

\begin{abstract}
Two, lightweight diode laser frequency stabilization systems designed for experiments in the field are described. A significant reduction in size and weight in both models supports the further miniaturization of measurement devices in the field.  Similar to a previous design, magnetic-field lines are contained within a magnetic shield enclosing permanent magnets and a  Rb cell, so that these DAVLL systems may be used for magnetically sensitive instruments. The Mini-DAVLL system (49~mm long) uses a vapor cell (20~mm long), and does not require cell heaters. An even smaller Micro-DAVLL system (9mm~long) uses a micro-fabricated cell (3~mm square), and requires heaters. These new systems show no degradation in performance with regard to previous designs, while considerably reducing dimensions.
\end{abstract}

\pacs{07.55.Ge, 32.60.+i, 42.65.-k, 85.70.Sq}
%07.55.Ge Magnetometers for magnetic field measurements
%32.60.+i Zeeman and Stark effects
%42.65.-k Nonlinear optics
%85.70.Sq Magneto-optical devices
\maketitle

\section{Introduction}
The development of laser frequency-locking systems has progressed in numerous directions as applications have called for greater precision and stability in optical frequency. A simple laser lock, the dichroic atomic-vapor laser lock (DAVLL) system \cite{1}, operates in Doppler-broadened mode, employing magnetic field induced circular dichroism and birefringence of the atomic vapor \cite{2}. The basic idea of the technique is that for an isolated atomic line, linear-optical circular dichroism induced by a longitudinal magnetic field has dispersive frequency dependence. The corresponding ellipticity induced in the transmitted light beam propagating along with the magnetic field turns to zero when the light is resonant with the atomic transition. A variant of the technique used to lock the frequency on the wings of an optical line uses linear-optical circular birefringence and the detection of optical rotation rather than induced ellipticity \cite {2}.

DAVLL systems can also be operated in the Doppler-free mode, in which a narrow differential absorption signal is used, resulting in a more precise locking of the laser frequency, at the cost of a smaller capture range \cite{3,4,5,6}. Doppler-broadened mode DAVLL systems are usually well-suited for the purposes of atomic magnetometry, and the DAVLL is commonly used for such experiments \cite{7,8}.

However, atomic magnetometers are also becoming more portable and compact to accommodate various applications from geophysical measurements, to biomagnetism \cite{9}, to the detection of nuclear magnetic resonance and magnetic-resonance imaging, see, for example, Ref. 10. As a result, there is a need for miniaturization of the associated instrumentation, including the system for locking the laser on an atomic resonance. Here, we report on realizations of DAVLL systems that are considerably reduced in size and weight compared with a previous design developed in our laboratory \cite{2}, which is cube-shaped, of dimensions $\sim$11.2 cm on each side, and of weight 6.5~kg. As in the case of that DAVLL system, magnetic field lines are completely contained within the magnetic shield enclosing the cell-magnet system, allowing operation in immediate proximity of sensitive optical magnetometry experiments. The first system presented in this paper (named Mini-DAVLL) uses a vapor cell of diameter 10~mm and length 20~mm, smaller than the one used in the DAVLL of Ref. 2 (20~mm diameter, 20~mm length).
\begin{figure}
\includegraphics[width=2.5 in]{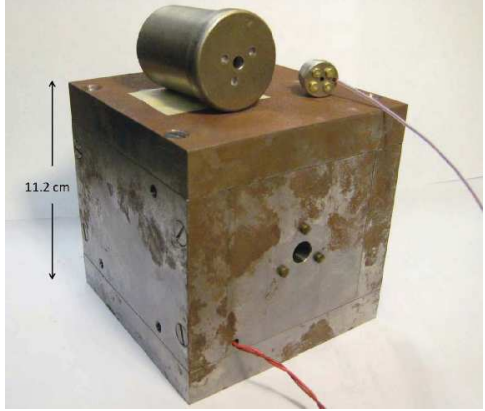}
\caption{(Color online) Size comparison: original Berkeley DAVLL design (bottom), Mini-DAVLL (top-left), and Micro-DAVLL (top-right).}
\label{BigDAVLL}
\end{figure}
Even though these cell sizes are comparable, the judicious realization of a uniform magnetic field, using a magnet configuration that surrounds the cell, allows for the drastic reduction in size. Thus, the basic conceptual design of the cell-magnet system of this new DAVLL is similar to that of Ref. 2, but the size of the entire system is greatly reduced (Fig. \ref{BigDAVLL}). The second DAVLL system presented (named Micro-DAVLL) uses a micro-fabricated vapor cell \cite{5,11}, which allows further significant reduction of the overall system size and weight, at the expense of requiring internal heaters. Both of these systems have potential for space borne applications, where size, weight, and power are essential considerations. A detailed description of the underlying theory and the optical schematic of the DAVLL system can be found in the previous publications of the JILA \cite{1} and Berkeley \cite{2} groups. In this article, we give a brief description of the cell-magnet system in the two newly developed DAVLL systems, and describe their optical characterization.

As in the case of the previous DAVLL model \cite{2}, the performance of these new DAVLL systems has been verified in sensitive magnetometry experiments. Previous experiments with these older models have found the laser drift over several minutes to be less than 1~MHz \cite{12}, which is within the appropriate range for magnetometry experiments \cite{1}. The systems in this report have been characterized with laser frequency drifts in the 100~KHz level over minute time scales (Figs. \ref{Minigraph} and \ref{micrographs}).

\section{Mini-DAVLL: The Cell-magnet System and Shield}
\begin{figure} [ht!]
\includegraphics[width=2.5 in]{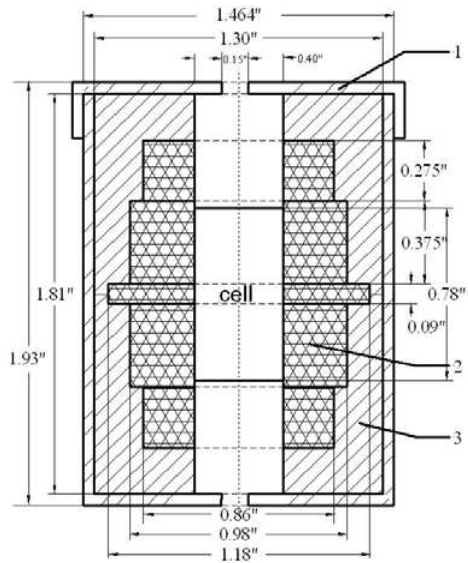}
\caption{Cross section of the Mini-DAVLL cell-magnet system. 1, CO-Netic (mu-metal) magnetic shield; 2, five flexible magnetic pole rings; 3, plastic support. The entire system is cylindrically symmetric about the axis indicated by the dashed line at the center of the figure.}
\label{MiniDrawing}
\end{figure}

\begin{figure} [ht!]
\includegraphics[width=2.4 in]{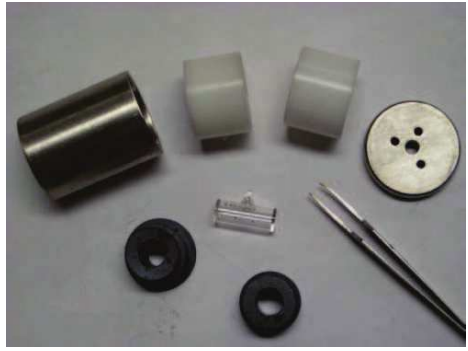}
\caption{(Color online) Mini-DAVLL parts - see Fig. \ref{MiniDrawing} for details}
\label{MiniParts}
\end{figure}

\begin{figure}
\includegraphics[width=2.4 in]{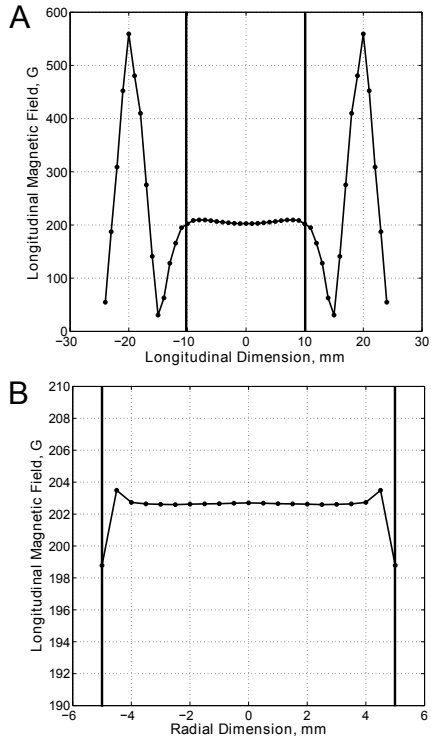}
\caption{Longitudinal magnetic field (along the light propagation axis) at the cell location: (A) dependence on longitudinal dimension (radial dimension is zero); (B) dependence on radial dimension (longitudinal dimension is zero). Bold lines represent the inner wall of the vapor cell. Values were calculated with COMSOL Multiphysics. Note the difference in vertical scales.}
\label{MiniField}
\end{figure}

The Mini-DAVLL utilizes the cylindrical shape of the $^{87}$Rb cell. Figures \ref{MiniDrawing} and \ref{MiniParts} show the cell-magnet system of the Mini-DAVLL. Cylindrical magnet poles, cut out from flexible permanent magnet sheets, have holes in the center to allow for the passage of a laser beam. A considerable amount of space was saved by placing these ring-shaped magnets around the cell, as opposed to placing them some distance away from the longitudinal ends of the cell. Magnet rings form a staircase pattern to maximize the uniformity of the magnetic field over the cell volume at 200~G, which is optimal for optical transitions, such as rubidium and cesium D lines \cite{2}. At this value of magnetic field, Zeeman splitting is comparable to the Doppler width and the DAVLL signal is maximal in size. COMSOL Multiphysics software was used to search for this ideal configuration given the distortion of the field due to the Co-Netic shielding (Fig. \ref{MiniField}). The magnets are held in place inside the cylindrical shield with a cylindrical plastic support. The total weight of the Mini-DAVLL is 0.18~kg.

Cell heaters are not used in the Mini-DAVLL system, as the appropriate absorption signal can be observed at room temperature. However, a simple modification to the white plastic magnet casing, as shown in Fig. \ref{MiniParts}, can be added to house heater wires in order to significantly amplify the signal and allow operation at low ambient temperatures. Soft-iron pucks \cite{2}, used in the previous Berkeley DAVLL, did not enhance the uniformity of the magnetic field in the Mini-DAVLL design.

The entire cell-magnet system is enclosed within a cylindrically shaped, 1.6~mm thick Co-Netic shield to minimize spillage of magnetic field outside the system. This Co-Netic shield consists of a cylindrical piece with one end open, and a cap that closely fits onto the cylinder, in a ``shoe-box" lid style.

\section{Performance of the Mini-DAVLL}
\begin{figure}[ht!]
\includegraphics[width=2.8 in]{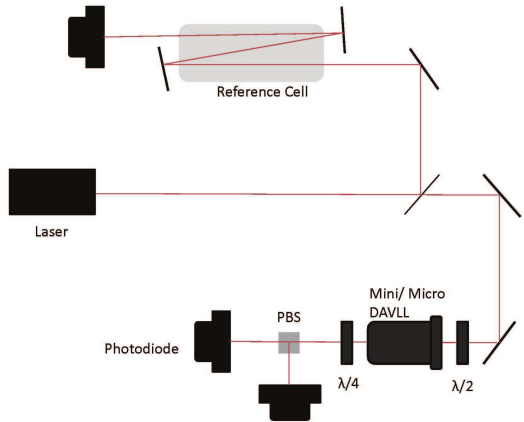}
\caption{(Color online) Optical schematic of experimental setup for characterizing the DAVLLs. PBS - polarizing beam splitter. Triple pass through reference cell is to improve absorption signal. Both modes of DAVLL operation can be achieved in this setup through a change in the relative axis of the wave plates.}
\label{minioptics}
\end{figure}

The experimental setup used to examine the performance of the Mini-DAVLL is shown in Fig. \ref{minioptics}. A small portion ($\sim0.5$ mW) of linearly polarized light from a 780 nm laser beam (corresponding to the Rubidium D2 transition) enters the DAVLL after passing through a half-wave plate, used to align the input polarization axis. On the tail end of the DAVLL, a quarter-wave plate is followed by a polarizing beam splitter (PBS). It was found that the polarization from the laser was sufficiently clean not to require an additional polarizer before the half-wave plate. With this setup, there are two possible modes of operation of the DAVLL.

The mode is selected by the relative positioning of the axes of the quarter-wave plate, input polarization, and output PBS. If the PBS is oriented at $\phi=45^{\circ}$ to the quarter-wave plate axes, the DAVLL is in the circular-analyzer mode (Mode 1), where the angle ($\alpha$) between the input light polarization and the PBS does not matter. In this mode, the characteristic DAVLL signal, corresponding to the vapor cell's circular dichroism, is observed. This configuration is used to lock the laser on
resonance. When the input light polarization is rotated to $\alpha=45^{\circ}$ by that half wave plate, and $\phi=0^{\circ}$, the DAVLL system is transformed into the balanced-polarimeter mode (Mode 2).

Here, a signal characteristic for the Macaluso-Corbino effect, corresponding to optical rotation, is observed. This second configuration can be used to tune the laser to the ``wings" of the resonance \cite{1}. Adjusting the axis of the quarter-wave plate continuously shifts the DAVLL operation between these two modes, allowing for a wide range of frequencies around resonance at which to lock the laser.

In examining the performance of both DAVLL systems presented in this paper, the laser was locked close to resonance through the use of the differential absorption signal of Mode 1. A proportional-integral-derivative (PID) controller interprets the differential absorption as an error signal and corrects it in a feedback system to the laser current driver. This error signal is also a reliable measure of frequency deviation from the desired locking point, with conversion factors calculated with respect to the absorption in a reference cell. This reference cell is set up along a separate section of the beam, with the laser triple-passed through it to improve the absorption signal (Fig. \ref{minioptics}).

\begin{figure}[ht!]
\includegraphics[width=2.6 in]{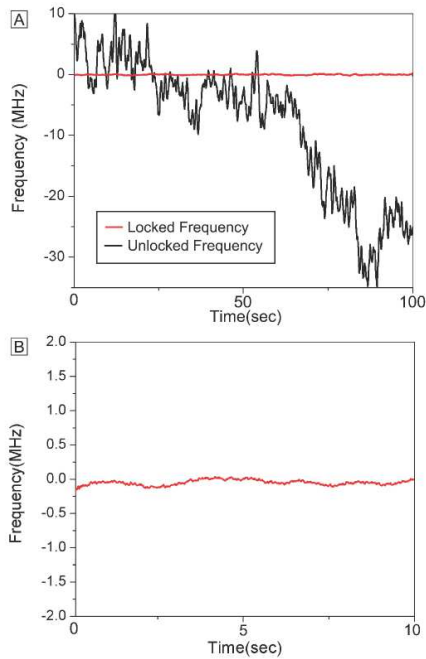}
\caption{(Color online) (A) Laser frequency drift of locked and unlocked laser over 100 second interval; (B) Mini-DAVLL locked laser frequency zoomed in on both axes over a 10 second interval; the running integration period is 1 second.}
\label{Minigraph}
\end{figure}

With the laser frequency unlocked over 100 seconds, the DAVLL signal fluctuated (RMS) in excess of 14~MHz from the reference point. With the PID on, this RMS value of the smoothed data dropped to 0.19~MHz over the entire time period. Deviation from the reference point is graphed in Fig. \ref{Minigraph}. To reduce electronic noise, data points over the time interval are smoothed in plotting software by taking a moving average, with a running integration period of 1 second.

\section{Micro-DAVLL: The Cell-magnet System and Shield}
\begin{figure} [ht!]
\includegraphics[width=5 in]{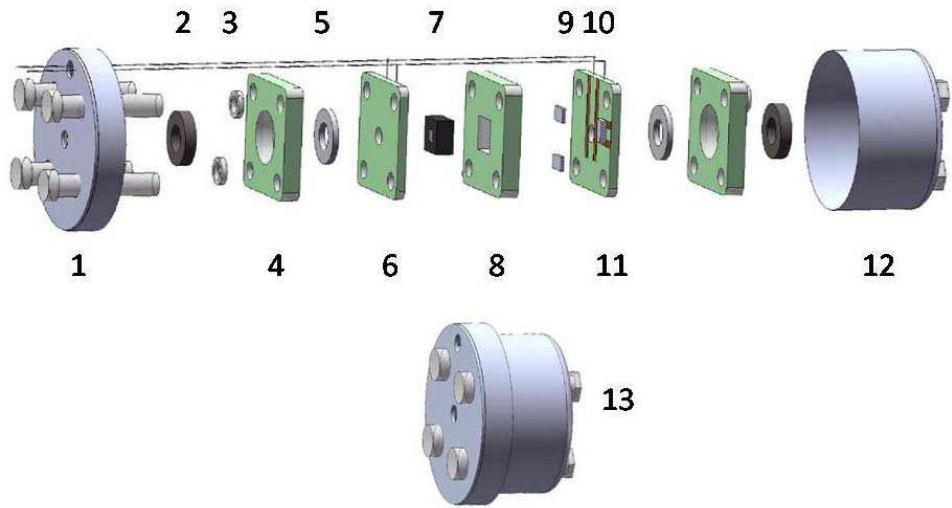}
\caption{(Color online) The cell-magnet system. 1, CO-Netic magnetic shield lid; 2, Flexible magnetic pole rings; 3, Plastic washers; 4, PC board housing magnets and pucks; 5, Co-Netic pole pucks; 6, Heater boards where cell heaters are soldered on; 7, Rubidium cell; 8, PC board housing the Rubidium cell; 9, 200 $\Omega$ resistors (cell heaters); 10, Wiring to heaters; 11, Thermistor soldered onto board (for clarity, wires to this unit are not shown); 12, Co-Netic magnetic shield; 13. Closed system. The system is held together by nylon screws running through the body.}
\label{MicroDrawing}
\end{figure}

The Micro-DAVLL has in its core a micro fabricated rubidium vapor cell developed by the NIST group \cite{11,13}. The cell is fabricated by bonding glass wafers to both sides of a silicon wafer. $^{87}$Rb atoms are contained in a cavity formed inside the silicon wafer (see Ref. \cite{11,13} for details). Figure \ref{MicroDrawing} shows the cell-magnet system of the Micro-DAVLL. As the size of the cell is much smaller than that of the Mini-DAVLL, the size and weight of the entire system could again be greatly reduced to the dimensions outlined below and weight of 0.65~g.

The Micro-DAVLL design is enclosed within a Co-Netic magnetic shield, with the lid fitting over the body in the same ``shoe-box" manner as in the Mini-DAVLL. This shield has dimensions 9~mm length, 14.5~mm diameter for the body, and 16~mm diameter for the lid (Fig. \ref{MicroDrawing}). Despite the smaller size of the cell, this first iteration of the Micro-DAVLL called for a larger shield relative to the size of the cell to maintain the homogeneity of the magnetic field over the volume of the cell. It was found that some inhomogeneity (\textless~20~G) in the field strength is acceptable with regard to performance, which is an important consideration for greater reduction of the shield size in subsequent iterations.

\begin{figure}
\includegraphics[width=5.6 in]{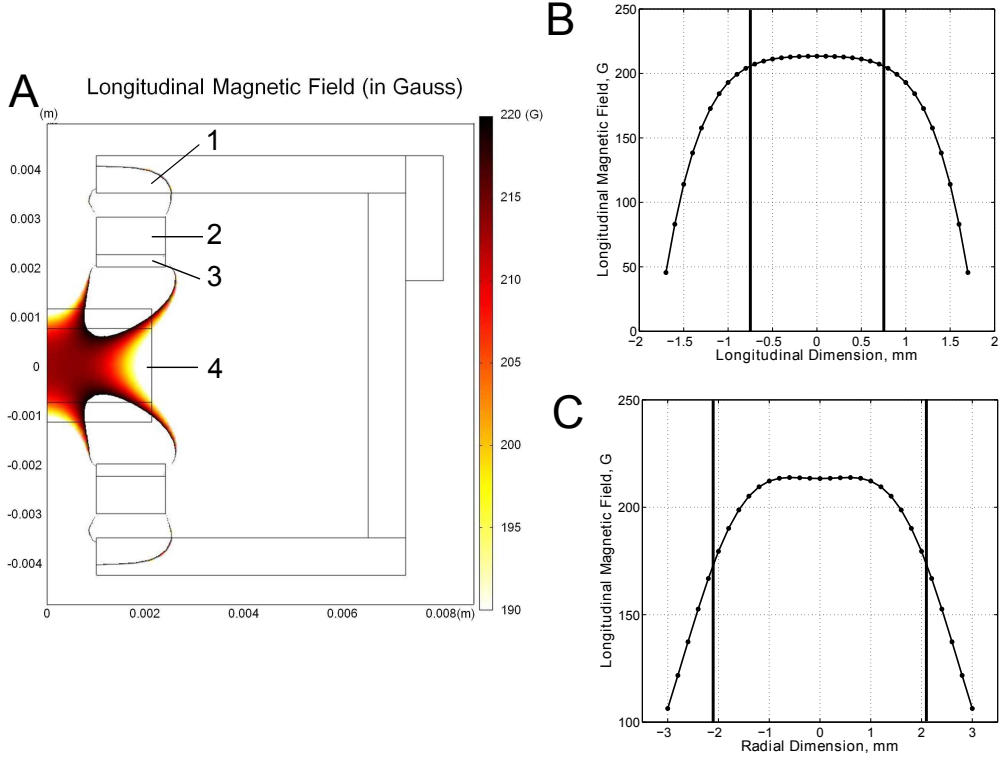}
\caption{(Color online) (A) Visualization of 200~G longitudinal component of magnetic field created in COMSOL Multiphysics. 1, Co-Netic Shield; 2, Magnet; 3, Pole Puck; 4, Cell. For clarity, other components of the DAVLL are not displayed. (B) and (C): Longitudinal magnetic field at the cell location; (B) dependence on longitudinal dimension (radial dimension is zero); (C) dependence on radial dimension (longitudinal dimension is zero). Bold lines represent the inner wall of the vapor cell. Values were calculated with COMSOL Multiphysics. Note the difference in vertical scales.}
\label{Magnetic Field}
\end{figure}

In the Micro-DAVLL, magnets are placed at the longitudinal ends of the cell, as this configuration is optimal for the flat and wide micro-fabricated cell. COMSOL Multiphysics was used to simulate the magnitude of the magnetic field over the cell, and to design the optimal configuration (Fig. \ref{Magnetic Field}). A pair of magnets and Co-Netic, soft iron pole pucks produce a uniform longitudinal magnetic field (200~G) at the cell. In the same manner as the cell, the magnets and pucks are held in place by blank printed-circuit (PC) boards.

\begin{figure}
\includegraphics[width=2.4 in]{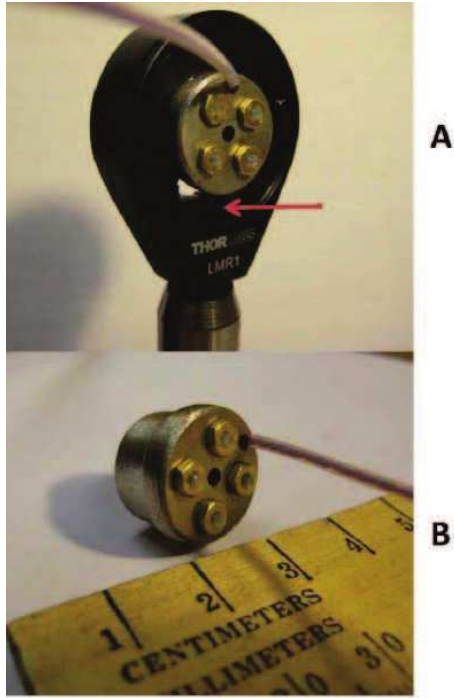}
\caption{(Color online) (A) Micro-DAVLL in Delrin mount (indicated by arrow), inside of Thor-Labs 1 in optical mount; (B) Closed Micro-DAVLL system}
\label{pics}
\end{figure}

Cell heaters are included in this model, since the amount of rubidium vapor in the micro-fabricated cell is not enough to generate a sufficient absorption signal at room temperature. To provide heat to the cell, two 200~$\Omega$, 1/8~W, surface mount resistors are soldered in parallel onto each of the two PC boards such that their magnetic field contributions cancel each other in the center of the cell (Fig. \ref{MicroDrawing}). A 5~V power supply is used to supply up to 100~mA of current to the heater system. A thermistor monitors temperature inside the cell. Depending on the application, this thermistor may also be used to actively stabilize the temperature by modulating the heater current. The system is mounted in a Delrin plastic mount that fits into a standard 1 in optical mount (Fig. \ref{pics}). With this Delrin mount and a lens holder, which minimizes contact points between the shield and the mount, internal temperature, as measured by the thermistor, was stabilized to $60 ^{\circ}$ C. At this temperature, one absorption length on the D2 resonance is observed.

The assembly is held together by use of nylon machine screws running through the case, secured by brass nuts, and the wires to the heaters lead out of the top of the case (Fig. \ref{MicroDrawing}). The layered plastic construction provides rigidity to the system, as well as insulation to maintain temperature at the cell. The cell is held in place by a blank PC board with a square hole in the center that allows for expansion of the cell as it is heated.

\section{Performance of the Micro-DAVLL}

\begin{figure} [ht!]
\includegraphics[width=2.6 in]{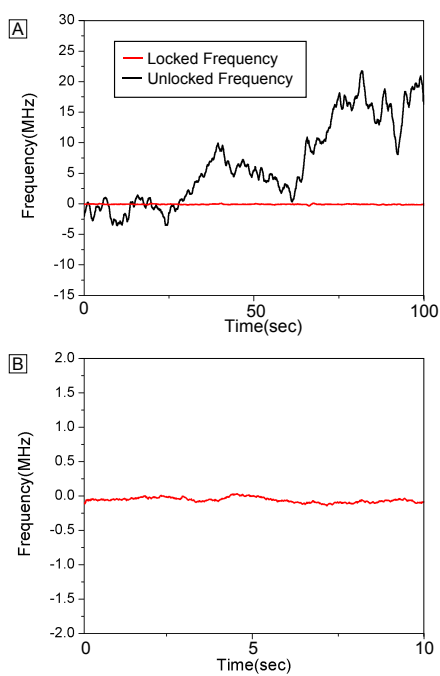}
\caption{(Color online) (A) Laser frequency drift of locked and unlocked laser over 100 second interval; (B) Micro-DAVLL locked laser frequency zoomed in over 10 second interval; running integration period is 1 second.}
\label{micrographs}
\end{figure}

The performance of the Micro-DAVLL system in stabilizing laser frequency is tested in the same manner as the Mini-DAVLL (Fig. \ref{minioptics}). With the laser frequency unlocked over 100 seconds, the DAVLL signal fluctuated (RMS) in excess of 16~MHz. With the PID on, this RMS value of the smoothed data dropped to 0.66~MHz over the same time period. Deviation from the reference point is graphed in Fig. \ref{micrographs}. As before, data points are smoothed by taking a moving average, with an effective integration period of 1 second.

\section{Smaller Systems}
In order to further reduce the size of the Micro-DAVLL system, it would be possible to arrange the heater wires in a solenoidal fashion around the inside radius of the shield so that the heaters would produce the required magnetic field, alleviating the need for permanent magnets \cite{14}. As a result, the shield size could be reduced by a factor of two or more, which is currently not possible because permanent magnets require a larger shield to maintain field homogeneity.

\begin{figure}
\includegraphics[width=3 in]{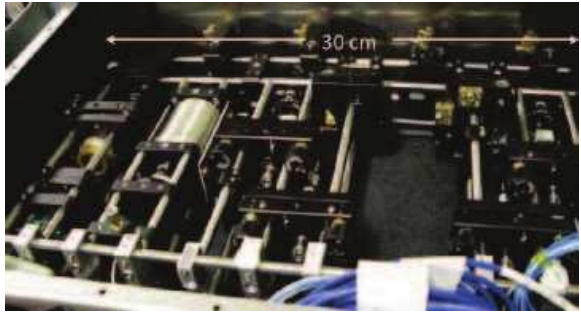}
\caption{(Color online) Application: Mini-DAVLL (silver/shiny cylinder) in an optical cage system of a UC Berkeley/Southwest Sciences portable magnetometer \cite{15}.}
\label{Cage_1}
\end{figure}

As the need for atomic clocks, magnetometers, and gyroscopes grows in fields where precise measurements are required, so will the need for these devices to become more portable and compact. Already, the DAVLL system has proven to be a reliable, compact component of a portable magnetometer \cite{15} (Fig. \ref{Cage_1}). The small-sized, lightweight DAVLL systems presented here are ideal for field applications where space and power supply are limited, in particular, air- and space-borne magnetometers.

\section{Acknowledgements}
The authors would like to sincerely thank J. Kitching and L. Garner for their useful discussions and advice on this work. This work was supported by the ONR MURI, and STTR programs, the NURI program, by the Director, Office of Science, Office of Basic Energy Sciences, Nuclear Science Divisions, of the U.S. Department of Energy under contract DE-AC03-76SF00098, and by the Microsystems Technology Office of the Defence Advanced Research Projects Agency (DARPA). This work is a partial contribution of NIST, an agency of the United States government, and is not subject to copyright. C. Lee and G. Iwata were participants in the U.C. Berkeley Undergraduate Research Apprentice Program.

Products or companies named here are cited only in the interest of complete technical description, and neither constitute nor imply endorsement by NIST or by the US government. Other products may be found to serve just as well.

\textit{Note added}: After completing this manuscript, we learned of related work by N. Kostinski \textit{et al.},  where the authors have analyzed the sensitivity of DAVLL systems to variation of vapor pressure \cite{16}.

\end{document}